\documentclass[preprint,12pt]{elsarticle}

\usepackage{graphicx}
\usepackage{amssymb}
\usepackage{amsthm}
\usepackage{amsmath}
\usepackage{makecell}
\usepackage{hyperref}

\newcounter{bla}

\journal{Computer Physics Communications}

\begin{document}

\begin{frontmatter}


\title{A Brief Introduction to PACIAE 4.0}

\author[a,b]{An-Ke Lei}
\author[c]{Zhi-Lei She}
\author[d]{Yu-Liang Yan}
\author[a]{Dai-Mei Zhou \corref{author}}
\author[e]{Liang Zheng}
\author[f]{Wen-Chao Zhang}
\author[f]{Hua Zheng}
\author[b]{Larissa V. Bravina}
\author[b,g]{Evgeny E. Zabrodin}
\author[a,d]{Ben-Hao Sa}

\cortext[author] {Corresponding author.\\\textit{E-mail address:} zhoudm@mail.ccnu.edu.cn}
\address[a]{Key Laboratory of Quark and Lepton Physics (MOE) and Institute of Particle Physics, Central China Normal University, Wuhan 430079, China}
\address[b]{Department of Physics, University of Oslo, P.O. Box 1048 Blindern, Oslo N-0316, Norway}
\address[c]{School of Mathematical and Physical Sciences, Wuhan Textile University, Wuhan 430200, China}
\address[d]{China Institute of Atomic Energy, P. O. Box 275 (10), Beijing 102413, China}
\address[e]{School of Mathematics and Physics, China University of Geosciences (Wuhan), Wuhan 430074, China}
\address[f]{School of Physics and Information Technology, Shaanxi Normal University, Xi'an 710119, China}
\address[g]{Skobeltsyn Institute of Nuclear Physics, Moscow State University, Vorob’evy Gory, Moscow RU-119991, Russia}

\begin{abstract}
Parton And-hadron China Institute of Atomic Energy (PACIAE) is a multipurpose 
Monte Carlo event generator developed to describe a wide range of high-energy 
collisions, including lepton-lepton, lepton-hadron, lepton-nucleus, 
hadron-hadron, hadron-nucleus, and nucleus-nucleus collisions. It is built 
based on the PYTHIA program, and incorporates parton and hadron cascades  
to address the nuclear medium effects. PACIAE 4.0 is the new generation of 
PACIAE model surpassing the version 3.0. In PACIAE 4.0, the old 
fixed-format FORTRAN 77 code has been refactored and rewritten by the 
free-format modern Fortran and C++ languages. The C++-based PYTHIA 8.3 is 
interfaced in, while previous versions connected to the Fortran-based PYTHIA 
6.4 only. Several improvements are also introduced, which enable PACIAE 4.0 to 
contain more physics and features to model the high-energy collisions. This is 
the first attempt to transition PACIAE from Fortran to C++.
\end{abstract}


\end{frontmatter}


{\bf PROGRAM SUMMARY}

\begin{small}
\noindent

{\em Program Title:} PACIAE 4.0 \\
{\em CPC Library link to program files:} (to be added by Technical Editor) \\
{\em Developer's repository link:} https://github.com/ArcsaberHep/PACIAE4 \\
{\em Code Ocean capsule:} (to be added by Technical Editor) \\
{\em Licensing provisions:} GPLv2 or later \\
{\em Programming language:} Fortran, C++ \\
{\em Supplementary material:}                                 \\
{\em Journal reference of previous version:} Computer Physics Communications 
284 (2023) 108615 \\
{\em Does the new version supersede the previous version?:} Yes \\
{\em Reasons for the new version:} improved and expanded physics models, 
transition from FORTRAN 77 to the modern Fortran mixed with C++ \\
{\em Summary of revisions:} PYTHIA 8 interface, transition to modern Fortran 
mixed with C++, and much more \\
{\em Nature of problem:} the Monte Carlo (MC) simulation has been successfully 
applied to the study of the high-energy collisions. MC models are able to give 
a fairly good description of the basic experimental observables. However, as 
more and more experimental data become available, more accurate modeling is 
required.\\
{\em Solution method:} the parton and hadron cascade model PACIAE 2 
series~[1-6] and 3.0~[7] are based on the Fortran-based PYTHIA 6.4~[8]. 
PYTHIA has been upgraded to the C++-based PYTHIA 8.3~[9] with more physics 
and features. Therefore we upgrade the PACIAE model to the new version of 
PACIAE 4.0 with the option to either PYTHIA 6.4~[8] or PYTHIA 8.3~[9]. 
In addition, several improvements have been introduced in this new version. \\
{\em Additional comments including restrictions and unusual features:} \\
Restrictions depend on the problem studied. The running time is 1–1000 
events per minute, depending on the collisions system studied. \\

\end{small}

\section{Introduction} \label{sec:introduction}
PACIAE is a parton and hadron cascade model to simulate the high-energy 
collisions. It stems from the LUnd and CIAE 
(LUCIAE) Monte Carlo program~\cite{Andersson:1996xi,Sa:1995fj,Tai:1998hc}, 
which connected to FRITIOF 7~\cite{Pi:1992ug,Andersson:1992iq}, 
JETSET 7~\cite{Sjostrand:1993yb}, PYTHIA 5~\cite{Sjostrand:1993yb} and 
ARIADNE 4~\cite{Lonnblad:1992tz}. LUCIAE implemented both the Firecracker Model 
(FCM) for the collective multigluon emission from the color fields of 
interacting strings and the reinteraction of the final state hadrons to study 
the collective and the rescattering effects. After that, the JETSET and PYTHIA 
CIAE (JPCIAE)~\cite{Sa:2001ma} was developed 
and detached FRITIOF and ARIADNE. Soon after, JETSET had been integrated into 
PYTHIA 6.1~\cite{Sjostrand:2000wi}. JPCIAE was renamed to PACIAE 
1.0~\cite{Sa:2004zc,Sa:2006zf,Zhou:2006ska} accordingly, and took the 
parton rescattering into account. In 2012, the first official version PACIAE 
2.0~\cite{Sa:2011ye} was released and then updated to version 
2.1~\cite{Sa:2012hb}, 
2.2~\cite{Zhou:2014qpa,Yan:2018mpp,She:2022vco,Yan:2022byw} and 
3.0~\cite{Lei:2023srp} over the past decade. However, in the previous PACIAE  
2 series~\cite{Sa:2011ye,Sa:2012hb,Zhou:2014qpa,Yan:2018mpp,She:2022vco,Yan:2022byw} 
and 3.0~\cite{Lei:2023srp}, the core component PYTHIA was restricted to the 
version 6.4~\cite{Sjostrand:2006za}, although it has been upgraded to the 
version 8~\cite{Sjostrand:2007gs,Sjostrand:2014zea,Bierlich:2022pfr}. A 
historical reason is that the older PACIAE and PYTHIA 6 were written with the 
process-oriented FORTRAN77 programming language, while PYTHIA 8 has embraced 
the objective-oriented C++. On the other hand, as recent experimental data and 
theoretical studies suggested~\cite{Weber:2018ddv,ALICE:2020msa,Vermunt:2019ecg,ALICE:2020wfu,ALICE:2021npz,Bierlich:2023okq}, 
more physical mechanisms, e.g. multiparton 
interactions (MPI), color reconnections (CR), etc. and the more accurate 
modeling are required. Therefore, we upgrade PACIAE 2 series and PACIAE 3.0 
to the version 4.0. In this version, the program is rewritten by the modern 
Fortran and C++ languages. Now PACIAE 4.0 is a Fortran-C++ mixing language 
program and is able to connect to the C++-based PYTHIA 8. At the same time, the 
excellent PYTHIA 6 is also retained to provide some convenient uses, such as 
the independent fragmentation (IF). The code structure of PACIAE 4.0 has been 
refactored and redesigned, which gives the possibility to use almost all 
features of PYTHIA. The mass-corrected cross sections of the heavy quarks 
related to parton-parton scatterings are introduced in the partonic 
rescattering stage. The coalescence hadronization model is improved, and 
additional hadron-hadron reaction channels are incorporated into the hadronic 
rescattering stage.
PACIAE 4.0 program is now hosted on the open source 
platforms GitHub~\footnote{https://github.com/ArcsaberHep/PACIAE4} and 
Gitee~\footnote{https://gitee.com/arcsaberhep/PACIAE4}.

\section{Program structure} \label{sec:strusture}

\subsection{Program flow} \label{subsec:program_flow}
\begin{figure}[htbp]
    \centering
    \includegraphics[width=1\textwidth]{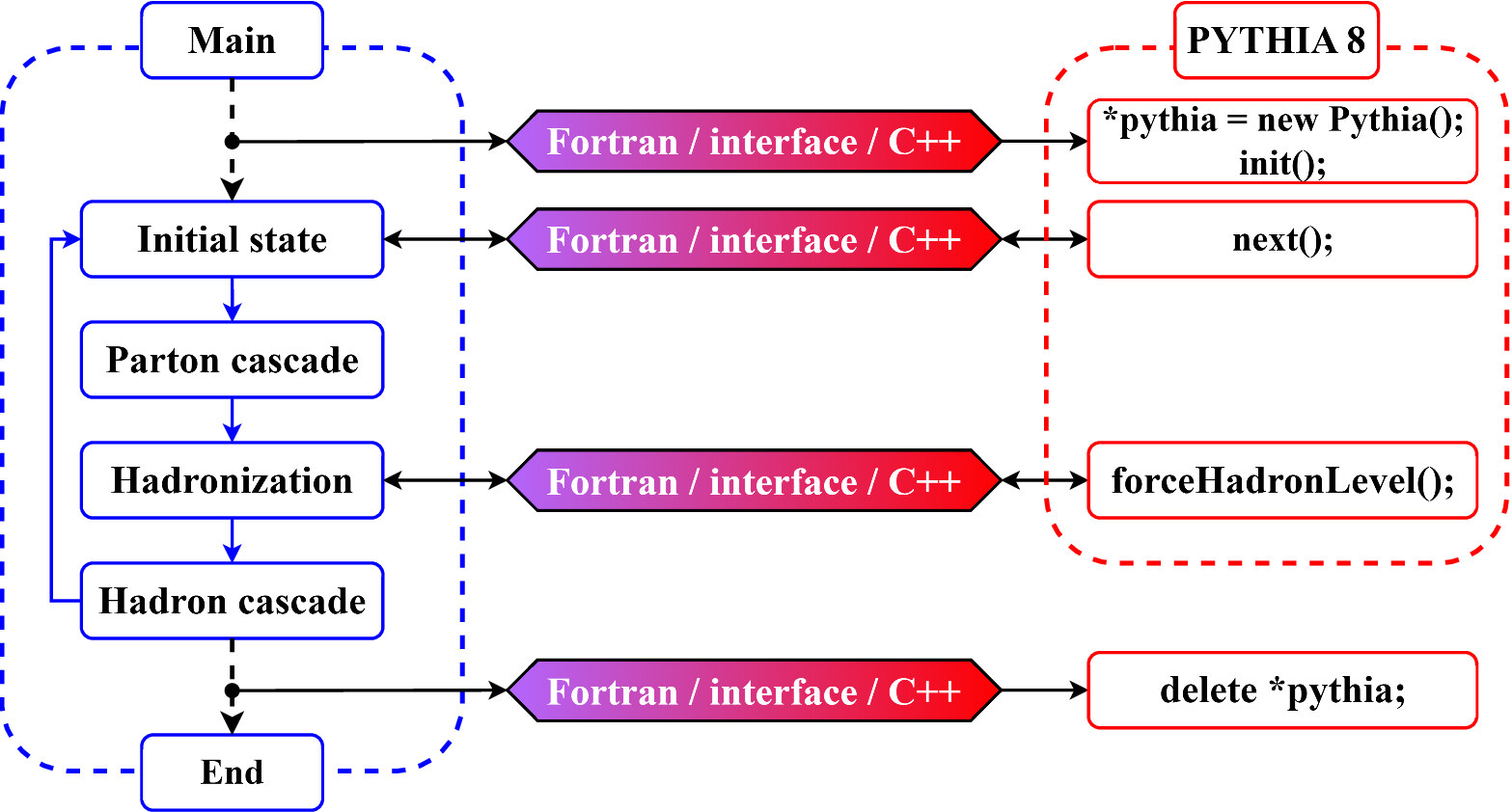}
    \caption{The program flow of PACIAE 4.0 when the PYTHIA 8 is chosen. 
             PACIAE interacts with C++-based PYTHIA 8 via the interface module.}
    \label{fig:program_flow}
\end{figure}

The program flow of PACIAE 4.0 is shown in Fig.~\ref{fig:program_flow}. In 
general, a full parton and hadron cascade simulation of PACIAE 4.0 comprise of 
four stages: initial state, parton cascade (rescattering), hadronization, 
and hadron cascade (rescattering). The skeleton of PACIAE 4.0 is still written 
in Fortran language. A Fortran main program steers the whole execution of 
PACIAE. At the beginning and the end, the main program instantiates and deletes 
PYTHIA 8 objects on the heap, respectively. For the initial state, PYTHIA 8 
provides the generation of hard processes, the construction of strings, etc. At 
the hadronization stage, if the Lund string fragmentation regime was selected, 
PYTHIA 8 will handle it and fragments parton strings to colorless hadrons. In 
these procedures, PACIAE interacts with PYTHIA 8 via the Fortran-C++ 
interface. Thanks to the INTERFACE block of the modern Fortran, it is easy 
to realize.

\subsection{Selective simulation frameworks} \label{subsec:framework}
The basic simulation frameworks are A-, B- and C-framework, 
corresponding to the low-energy ($\sqrt{s_{NN}} < 3 $ GeV) simulation, 
the high-energy ($\sqrt{s_{NN}} \ge 3 $ GeV) hadronic simulation and the 
high-energy parton and hadron cascade simulation inherited from PACIAE 
3.0~\cite{Lei:2023srp}. In PACIAE 4.0, according to the different initial state 
provider, we design several different frameworks (modes):
\begin{itemize}
    \item[(1)] low-energy hadronic simulation (A).
    \item[(2)] high-energy hadronic simulation based on PYTHIA 6 (B\_PY6).
    \item[(3)] high-energy parton and hadron cascade simulation based on 
               PYTHIA 6 (C\_PY6).
    \item[(4)] high-energy hadronic simulation based on PYTHIA 8 (B\_PY8).
    \item[(5)] high-energy parton and hadron cascade simulation based on 
               PYTHIA 8 (C\_PY8).
    \item[(6)] high-energy hadronic simulation based on PYTHIA8/Angantyr 
               (B\_ANG).
    \item[(7)] high-energy parton and hadron cascade simulation based on 
               PYTHIA8/Angantyr (C\_ANG).
\end{itemize}

In the following sections, we will focus on the parton and hadron cascade 
process. For the purely hadronic simulation in PACIAE, we refer to 
Ref.~\cite{Lei:2023srp} for more details.

\subsection{Input files} \label{subsec:input_file}
The parameters and switches controlling the simulation can be divided into 
two categories: PACIAE-related and PYTHIA-related. In the previous versions of 
PACIAE, all of the input parameters, both PACIAE-related and PYTHIA-related, 
were set in a single ``usu.dat'' file. In PACIAE 4.0, we decouple these two 
sets into an input file ``usu.dat'' for the main PACIAE-related parameters and 
two extra input files ``pythia6\_extra.cfg'' and ``pythia8\_extra.cfg'' for the 
PYTHIA6/8-related parameters
\footnote{ To use the ``pythia6\_extra.cfg'' and 
``pythia8\_extra.cfg'', the knowledge of PYTHIA parameters and switches is 
required. Important references include: 
(1) \href{https://doi.org/10.1088/1126-6708/2006/05/026}{PYTHIA 6.4 Physics 
and Manual, JHEP 05 (2006) 026}; 
(2) \href{https://pythia.org/download/pythia6/pythia6428.update}{PYTHIA 6.4 
Update Notes}; (3) \href{https://pythia.org//latest-manual/Welcome.html}{PYTHIA 
8 Online Manual, https://pythia.org//latest-manual/Welcome.html}. }. 
Essentially, ``pythia6\_extra.cfg'' and 
``pythia8\_extra.cfg'' files are  the ``card files'' of PYTHIA 6 and 
8~\cite{Sjostrand:2006za,Bierlich:2022pfr}. They allow more flexibility 
and freedom in the use of PYTHIA 6 and 8.

\section{Physics aspects} \label{sec:physics}

\subsection{Initial state} \label{subsec:initial_state}
\subsubsection{$pp$ collisions} \label{subsec:pp_collisions}
A proton-proton ($ pp $) pair is firstly executed by PYTHIA 8. PYTHIA 8 generates 
the hard processes at the leading order (LO), the space-like initial-state 
radiation (ISR) and the time-like final-state radiation (FSR) with the parton 
shower approach, as well as the MPI in the interleaved 
evolution~\cite{Bierlich:2022pfr,Corke:2010yf}. The strings from PYTHIA 
including hard, semi-hard, soft partons and the beam remnants (diquarks mainly) 
are extracted. Then the strings and diquarks are forced to break up into free 
partons to obtain the initial partonic state of the $ pp $ collision.

\subsubsection{$pA$ and $AB$ collisions} \label{subsec:pA_AB_collisions}
The positions of nucleons inside a nucleus $ A $ are sampled with the 
Woods-Saxon distribution
\begin{equation}
    \rho(r) = \rho_0 [ 1 + \exp{ ( \frac{ r - R }{d} ) } ]^{-1}, 
    \label{eq:woods_saxon}
\end{equation}
where $ r $ is the distance between a nucleon and the center of the nucleus, 
$ R $ is the classical radius of the nucleus, and $ d \approx 0.54 $ fm is the 
nuclear depth. The normalization constant $ \rho_0 $ reads
\begin{equation}
    \rho_0 = \frac{ A }{ \frac{ 4 \pi }{ 3 } R^3 }. 
    \label{eq:woods_saxon_norm_rho0}
\end{equation}
Here we use $ A $ denoting both the nucleus and its atomic number.
$ R $ can 
be expressed by the formula
\begin{equation}
    R = r_0 A^{1/3}, 
    \label{eq:A_radius}
\end{equation}
where $ r_0 = 1.12 $ fm is a radius parameter.

The beam direction is set along $ z $ axis. For $ pA $  and $ AB $ collisions, 
the projectile and the target are positioned at $ ( +\frac{b}{2}, 0, -20 )$ and 
$ ( -\frac{b}{2}, 0, +20 )$ (in units of fm), respectively. $ b $ is the impact 
parameter between the projectile and the target. The impact parameter 
angle (event-plane angle) $ \phi_{ b } $ 
is fixed to 0. In PACIAE 4.0, the upper limit of the nucleus radius is set as 
10 fm $ > R$, which means the maximum impact parameter will be 20 fm, similar 
to the Monte Carlo Glauber (MCG) 
model~\cite{Alver:2008aq,Loizides:2014vua,Loizides:2017ack}. 
It enables us to capture the nucleon fluctuation in the peripheral 
collisions. The time origin is set to the moment when the first nucleon-nucleon 
($ NN $) collides. Due to the very high collision energy, the Fermi motion 
of nucleons is neglected, i.e., the transverse momenta $ p_x = p_y = 0 $. 
The longitudinal momentum $ p_z $ and the energy $E$ of nucleons will be 
assigned according to the required collision energy. 
Two nucleons will collide if their minimum distance $ d_{min} $ satisfies 
\begin{equation}
    d_{min} \le \sqrt{ \frac{ \sigma^{NN}_{tot} }{ \pi } },
    \label{eq:d_min}
\end{equation}
where $ \sigma^{NN}_{tot} $ is the total $ NN $ cross section in units of fm$^2$ 
and $ d_{min} $ is defined in the center-of-mass (CM) frame of the two 
colliding nucleons. This criterion is also applied to the partonic 
and hadronic rescattering stages, just with different particle species and 
total cross sections. The $ \sigma^{NN}_{tot} $ is calculated as the sum of the 
pomeron and the reggeon terms using the widely-accepted Donnachie-Landshoff 
(DL) parameterization~\cite{Donnachie:1992ny} 
\begin{equation}
    \sigma^{NN}_{\mathrm{tot}}(s) = X s^{\epsilon} + Y s^{-\eta} ~,
\label{pg:sigtotpomreg}
\end{equation}
where $ s $ is the squared CM energy. The powers are $\epsilon = 0.0808$ and
$\eta = 0.4525$. The coefficients $ X $ and $ Y $ are 21.7 and 56.08, 
respectively.

Each colliding $ NN $ pair will be executed by PYTHIA 8, and free partons are 
obtained in the manner described in Sec.~\ref{subsec:pp_collisions}. 
These partons compose the initial partonic state of a $ pA $ or $ AB $ 
collision.

\subsubsection{Angantyr initial state} \label{subsec:angantyr}
With the upgrade of PYTHIA, the heavy-ion simulations, the Angantyr 
model~\cite{Bierlich:2018xfw}, has been developed and integrated into the 
standard PYTHIA 8 package~\cite{Bierlich:2022pfr}. Angantyr is inspired by the 
old FRITIOF model~\cite{Pi:1992ug,Andersson:1992iq,Nilsson-Almqvist:1986ast} 
and inherits the name from one of its subroutines directly. Compared to the  
standard Glauber formalism, Angantyr implements the more advanced GLISSANDO 
model~\cite{Broniowski:2007nz,Rybczynski:2013yba} and the DoubleStrikman 
model~\cite{Alvioli:2013vk}, in which the elaborate color fluctuations inside 
nucleons are coded. Recent studies~\cite{Zheng:2021jrr,Zheng:2024xyv} have 
shown such a substructure might be of importance to the collectivity of the small 
systems. In PACIAE 4.0, it is able to generate an Angantyr initial 
partonic state following the same way in Sec.~\ref{subsec:pp_collisions}, then 
push the parton reservoir into the parton cascade evolution.

\subsubsection{Special pure quarks-antiquarks initial state} \label{subsubsec:quark_ini_state}
While the aforementioned initial states consist of the relatively large number of 
jet partons, PACIAE could provide another special initial state: deexcited 
quarks-antiquarks state. In this special initial state, there are only 
quarks and antiquarks. The gluons are converted to quark-antiquark ($q \bar{q}$) 
pairs, and the energetic quarks/antiquarks excite $q \bar{q}$ pairs from the 
vacuum. The details are described in Sec.~\ref{subsubsec:coalescence}.

\subsubsection{Other beams} \label{subsubsec:other_beam}
Since PACIAE is built based on the PYTHIA program, in addition to the basic 
$ pp / pA / AB $, it can accept any beams 
supported by PYTHIA 6/8, including but not limited to the electron $ e^{-} $, 
electron neutrino $ \nu_e $, photon $ \gamma $, pion $ \pi^+ $, 
antiproton $ \bar{p} $, $ \Lambda $ hyperon, etc. 
We refer to Refs.~\cite{Zhou:2014qpa,Sjostrand:2006za,Bierlich:2022pfr} for 
more information.

\subsection{Parton cascade} \label{subsec:parton_cascade}

\subsubsection{Parton-parton rescattering} \label{subsubsec:parton_rescattering}
The possible signals for the formation of the hot and dense nuclear 
matter, quark gluon plasma (QGP), have been detected by the Relativistic Heavy 
Ion Collider (RHIC)~\cite{BRAHMS:2004adc,PHOBOS:2004zne,STAR:2005gfr,PHENIX:2004vcz} 
and the Large Hadron Collider (LHC)~\cite{ALICE:2010suc,ALICE:2010yje,ATLAS:2010isq,CMS:2011iwn}. 
To study such a unique QCD matter, it is inevitable to model the interactions 
between deconfined partons. The Monte Carlo technique is a very effective 
method~\cite{Geiger:1991nj}, and there are copious parton transport 
models in the market, such as ZPC (AMPT)~\cite{Zhang:1997ej,Lin:2004en}, 
PCPC~\cite{Borchers:2000wf,Nara:2023vrq}
BAMPS~\cite{Xu:2004mz}, PHSD~\cite{Cassing:2009vt}, 
MARTINI~\cite{Schenke:2009gb}, PACIAE~\cite{Sa:2011ye}, 
ALPACA~\cite{Kurkela:2022qhn}, etc. In PACIAE, we consider only 
$ 2 \rightarrow 2 $ parton-parton scatterings ( $ a b \rightarrow c d $ ) with 
the leading-order perturbative quantum chromodynamics (LO-pQCD) cross 
sections~\cite{Combridge:1977dm}. The differential cross section reads 
\begin{equation}
    \frac{ d\sigma }{ dt } ( ab \rightarrow cd; s, t ) =
    \frac{ \pi \alpha_s^2 }{ s^2 } | \overline{ \mathcal{ M } } |^2,
\end{equation}
where $ s $ and $ t $ are squared CM energy and squared momentum transfer, 
respectively. $ \mathcal{ M } $ is the matrix elements. 
As a good approximation at the high-energy limit, in old versions of PACIAE, 
quarks (antiquarks) are treated as massless. Thus one has the relation 
$ s + t + u = 0 $ of the Mandelstam variables. The involved processes and 
corresponding matrix elements are listed in Table~\ref{tab:massless_process}, 
where `1' and `2' indicate different flavors.
In the listed matrix elements, the second forms are obtained by the small angle 
approximation.
\begin{table}[htbp]
    \caption{Matrix elements for the massless $ 2 \rightarrow 2 $ 
             parton-parton scattering processes.}
    \setlength{\tabcolsep}{0.06\textwidth}
    \renewcommand{\arraystretch}{2}
    \begin{tabular}{ c | c | c }
    \hline
    Order & Process & $ | \overline{ \mathcal{ M } } |^2 $ \\
    \hline
    1 & \makecell{ $ q_1 q_2 \rightarrow q_1 q_2 $ \\
                   $ \bar{ q_1 } \bar{ q_2 } 
                   \rightarrow \bar{ q_1 } \bar{ q_2 } $ } & 
        $ \frac{4}{9} \frac{ s^2 + u^2 }{ t^2 } 
        \approx \frac{8}{9} \frac{ s^2 }{ t^2 } $ \\
    \hline
    2 & $ q_1 q_1 \rightarrow q_1 q_1 $ & 
        $ \frac{4}{9} ( \frac{ s^2 + u^2 }{ t^2 } 
        + \frac{ s^2 + t^2 }{ u^2 }) - \frac{8}{27} \frac{ s^2 }{ ut } 
        \approx \frac{8}{9} \frac{ s^2 }{ t^2 } $ \\
    \hline
    3 & $ q_1 \bar{q_2} \rightarrow q_1 \bar{ q_2} $ & 
        $ \frac{4}{9} \frac{ s^2 + u^2 }{ t^2 } 
        \approx \frac{8}{9} \frac{ s^2 }{ t^2 } $ \\
    \hline
    4 & $ q_1 \bar{ q_1 } \rightarrow q_2 \bar{ q_2 } $ & 
        $ \frac{4}{9} \frac{ t^2 + u^2 }{ s^2 } 
        = \frac{4}{9} \frac{ t^2 + (t + s)^2 }{ s^2 } $ \\
    \hline
    5 & $ q_1 \bar{ q_1 } \rightarrow q_1 \bar{ q_1 } $ & 
        $ \frac{4}{9} ( \frac{ s^2 + u^2 }{ t^2 } 
        + \frac{ t^2 + u^2 }{ s^2 } ) - \frac{8}{27} \frac{ u^2 }{ t s } 
        \approx \frac{8}{9} \frac{ s^2 }{ t^2 }$ \\
    \hline
    6 & $ q \bar{ q } \rightarrow g g $ & 
        $ \frac{32}{27} \frac{ u^2 + t^2}{ u t } 
        - \frac{8}{3} \frac{ u^2 + t^2 }{ s^2 } 
        \approx -\frac{32}{27} \frac{ s }{ t } $ \\
    \hline
    7 & $ g g \rightarrow q \bar{ q }$ & 
        $ \frac{1}{6} \frac{ u^2 + t^2 }{ u t } 
        - \frac{3}{8} \frac{ u^2 + t^2 }{ s^2 } 
        \approx -\frac{1}{3} \frac{ s }{ t } $ \\
    \hline
    8 & \makecell{ $ q g \rightarrow q g $\\
                   $ \bar{q} g \rightarrow \bar{q} g $ } & 
        $ - \frac{4}{9} \frac{ u^2 + s^2 }{ u s } + \frac{ u^2 + s^2 }{ t^2 } 
        \approx \frac{ 2 s^2 }{ t^2 } $ \\
    \hline
    9 & $ g g \rightarrow g g $ & 
        $ \frac{9}{2} ( 3 - \frac{ u t }{ s^2 } 
        - \frac{ u s }{ t^2 } - \frac{ s t }{ u^2 } ) 
        \approx \frac{9}{2} \frac{ s^2 }{ t^2 } $ \\
    \hline
    \end{tabular}
    \label{tab:massless_process}
\end{table}

The corresponding integral cross section can be calculated as
\begin{equation}
    \sigma( a b \rightarrow c d; s ) =
        \int_{ -s }^0 \frac{ d\sigma }{ dt } ( a b \rightarrow c d; s, t ) dt.
    \label{eq:integral_xsect}
\end{equation}
In addition, the singularities appearing at $ t \rightarrow 0 $ can be 
regulated by introducing a cutoff of the Debye screening coefficient 
$ \mu $~\cite{Lei:2023srp,Zhang:1997ej}. Mathematically, the integral cross 
section evaluated using the cutoff regulator is equivalent to changing the 
lower and upper integral limits in Eq.~(\ref{eq:integral_xsect}):
\begin{equation}
    \sigma( a b \rightarrow c d; s ) =
        \int_{ -s - \mu^2 }^{ - \mu^2 } 
        \frac{ d\sigma }{ dt } ( a b \rightarrow c d; s, t ) dt.
    \label{eq:integral_xsect_mu}
\end{equation}

Furthermore, to expand our model to the heavy quark sector and 
achieve more accurate simulation, we consider the mass correction to the cross 
sections of heavy quarks in PACIAE 4.0. The relevant matrix elements 
are~\cite{Combridge:1978kx}:

\noindent 10. $ Q q \rightarrow Q q $
\begin{equation}
    | \overline{ \mathcal{ M } } |^2 = \frac{4}{9} 
    \frac{ ( M^2 - u )^2 + ( s - M^2 )^2 + 2 M^2 t }{ t^2 }; 
    \label{eq:ME_Qq}
\end{equation}

\noindent 11. $ Q g \rightarrow Q g $
\begin{equation}
    \begin{aligned}
        | \overline{ \mathcal{ M } } |^2 
            &= 
            \frac{ 2 ( s - M^2 ) ( M^2 - u )}{ t^2 } 
            + \frac{4}{9} \frac{ (s-M^2 ) ( M^2 - u ) 
            + 2 M^2 ( s + M^2 )}{ ( s - M^2 )^2 } \\
            &+ \frac{4}{9} \frac{ ( s - M^2 ) ( M^2 - u ) 
            + 2 M^2 ( M^2 + u ) }{ ( M^2 - u )^2 } 
            + \frac{1}{9} \frac{ M^2 (4 M^2 - t) }{ ( s - M^2 )( M^2 - u ) } \\
            &+ \frac{ ( s - M^2 )( M^2 - u ) + M^2( s - u ) }{ t ( s - M^2 ) } 
            - \frac{ ( s - M^2 ) ( M^2 - u ) - M^2( s - u ) }{ t ( M^2 - u ) } .
    \end{aligned}
    \label{eq:ME_Qg}
\end{equation}
In Eqs.~(\ref{eq:ME_Qq}) and (\ref{eq:ME_Qg}), $ M $ denotes the mass of 
heavy quark $c$ or $b$. We neglect the scatterings between two heavy quarks and 
the thermal production of heavy quarks at present.

\subsubsection{Medium-induced radiation} \label{subsubsec:parton_shower}
When jet partons traverse the medium, they might lose energies, i.e. so-called 
jet quenching phenomenon~\cite{Bjorken:1982tu,Gyulassy:1990ye}. The part due to 
the collisions between jet partons and medium partons is denoted as the 
collisional energy loss. We encode it in the parton-parton rescattering as 
described in Sec.~\ref{subsubsec:parton_rescattering}. 
Another important source of the energy loss comes from the medium-induced 
radiation, i.e. radiative energy loss. In PACIAE, we implement preliminarily 
the medium-induced radiation using the time-like shower approach. The simple 
collinear parton splitting~\cite{Webber:1986mc} is employed. We assume the 
outgoing final states from a parton-parton rescattering would radiate one or 
more partons with QCD branchings $ q \rightarrow q g $, $ g \rightarrow g g $ 
and $ g \rightarrow q \bar{q} $. The non-branching probability is determined by 
the Sudakov form factor~\cite{Sudakov:1954sw}. The Altarelli–Parisi (AP) 
splitting functions~\cite{Altarelli:1977zs} are utilized as splitting kernels. 
Whereas the collinear and infrared singularities are regularized by some scale 
cutoffs. Because of the sophistication, it is still under development. We leave 
this part for the next work.

\subsection{Hadronization} \label{subsec:hadronization}
PACIAE 4.0 provides two different hadronization mechanisms: the string 
fragmentation model and the coalescence model. Recently, the hybrid 
hadronization mechanisms of fragmentation + coalescence~\cite{Minissale:2020bif,Plumari:2017ntm,Song:2015sfa,Song:2015ykw,He:2011qa,He:2019vgs,Zhao:2023nrz}, 
even hydro-freeze-out + fragmentation + coalescence~\cite{Zhao:2020wcd,YuanyuanWang:2023wbz} 
have been progressively proposed and successfully predict the experimental data, 
especially in the heavy quark sector. These mechanisms have not yet been 
considered in PACIAE. Still, we leave them to the next version. In the present 
version, the fragmentation and the coalescence are separated.

\subsubsection{String fragmentation model} \label{subsubsec:fragmentation}
After the parton cascade evolution, the diquarks will be recovered in order to 
preserve the basic baryon number conservation. Partons are also recovered back 
to the strings (including the junction structure~\cite{Sjostrand:2006za,Bierlich:2022pfr}, 
a complicated string topology and another source of the baryon number 
conservation). Then the parton strings will be fed back to PYTHIA and be 
hadronized into colorless hadrons using the Lund string fragmentation regime.

\subsubsection{Coalescence model} \label{subsubsec:coalescence}
An alternative is the coalescence hadronization model. As a first step, PACIAE 
will split gluons to $q \bar{q}$ pairs. In the improved coalescence 
model, a gluon is assumed to be 
split to a $q \bar{q}$ pair sharing the forward light-cone momentum 
$ E + p_z $ of the gluon according to the AP splitting 
function~\cite{Altarelli:1977zs}
\begin{equation}
    f(z)_{ g \rightarrow q \bar{ q } } \propto z^2 + (1 - z)^2,
    \label{AP_split_function}
\end{equation}
where $ z $ and $ 1 - z $ are the light-cone momentum fraction taken by the 
quark and the antiquark from the mother gluon, respectively. Additionally, the 
split $ q $ and $ \bar{q} $ acquire locally compensated transverse momenta 
\begin{equation}
\begin{aligned}
    p_{x}^{q} = -p_{x}^{ \bar{q} }, \\
    p_{y}^{q} = -p_{y}^{ \bar{q} },
\end{aligned}
\end{equation}
differing from the collinear splitting. $ p_x $ and $ p_y $ are sampled from 
independent Gaussian distributions separately. A further treatment is the 
energetic quark (antiquark) deexcitation. We assume an energetic quark 
(antiquark) excites new $ q \bar{ q } $ pairs from the vacuum iteratively. 
The transverse momenta of $ q $ and $ \bar{q} $ are also sampled from Gaussian 
distributions compensatively. For the longitudinal direction, the 
$ q \bar{ q } $ takes a fraction of the forward light-cone momentum from its 
mother quark (antiquark) $ q_0 $ according to a 
Lund-form~\cite{Sjostrand:2006za} deexcitation function
\begin{equation}
    f(z)_{q_0 \rightarrow q\bar{q} + q_0} 
    \propto z^{-1}(1 - z)^a \exp{ (-b m_T^2 / z) },
\end{equation}
where $ a $ and $ b $ are two free parameters. 
$ m_T^2 = m^2 + p_T^2 $ is the squared transverse mass of the $ q \bar{ q } $ 
object with its mass and transverse momenta defined as
\begin{equation}
    \begin{aligned}
    m   &= m_q + m_{ \bar{ q } }, \\
    p_x &= p_x^q + p_x^{ \bar{ q } }, \\
    p_y &= p_y^q + p_y^{ \bar{ q } }.
    \end{aligned}
\end{equation}
It is worth mentioning that these treatments can be performed to obtain a 
special quarks-antiquarks initial state before the parton cascade.

The quarks and antiquarks then coalesce into hadrons using a simple 
phenomenological coalescence model~\cite{Lei:2023srp}. One can demand the 
phase-space constraint
\begin{equation}
    \frac{ 16 \pi^2 }{ 9 } (\Delta r)^3 (\Delta p)^3 = \frac{ h^3 }{ d },
\end{equation}
where $h^3/d$ is the volume occupied by a single hadron in the phase space, 
$d$=4 refers to the spin and parity degeneracies of the hadron. $\Delta r$ 
and $\Delta p$ stand for the sum of pair-wise relative distances between two 
(meson) or among three (baryon) partons in the spatial and momentum phase 
spaces, respectively. For partons that fail to coalesce with the 
constraint, they will be forced to coalesce into hadrons without phase-space 
constraint in order to satisfy the flavor conservation at the end of 
coalescence.

\subsection{Hadron cascade} \label{subsec:hadron_cascade}

\subsubsection{Hadron-hadron rescattering} \label{subsubsec:hadron_rescattering}
PACIAE 4.0 carries hundred of inelastic reaction channels of light hadrons 
developed from the original LUCIAE~\cite{Andersson:1996xi,Sa:1995fj,Tai:1998hc}
and old versions of 
PACIAE~\cite{Sa:2011ye,Sa:2012hb,Zhou:2014qpa,Yan:2018mpp,She:2022vco,Yan:2022byw,Lei:2023srp}. 
In this new version, we include more inelastic reactions involving the heavy 
hadrons: 
\begin{equation*}
    \begin{aligned}
        \pi N          &\rightleftharpoons  \pi \Delta,            &\quad
        \pi N          &\rightleftharpoons  \rho N,                \\
        N N            &\rightleftharpoons  N \Delta,              &\quad
        \pi \pi        &\rightleftharpoons  K \bar{K},             \\
        \pi N          &\rightleftharpoons  K Y,                   &\quad
        \pi \bar{N}    &\rightleftharpoons  \bar{K} \bar{Y},       \\
        \pi Y          &\rightleftharpoons  K \Xi,                 &\quad
        \pi \bar Y     &\rightleftharpoons  \bar{K} \bar{\Xi},     \\
        \bar K N       &\rightleftharpoons  \pi Y,                 &\quad
        K \bar{N}      &\rightleftharpoons  \pi \bar{Y},           \\
        \bar{K} Y      &\rightleftharpoons  \pi \Xi,               &\quad
        K \bar{Y}      &\rightleftharpoons  \pi \bar{\Xi},         \\
        \bar{K} N      &\rightleftharpoons  K \Xi,                 &\quad
        K \bar{N}      &\rightleftharpoons  \bar{K} \bar{\Xi},     \\
        \pi \Xi        &\rightleftharpoons  K \Omega,              &\quad
        \pi \bar{\Xi}  &\rightleftharpoons  \bar{K} \bar{\Omega},  \\
        K \bar{\Xi}    &\rightleftharpoons  \pi \bar{\Omega},      &\quad
        \bar{K} \Xi    &\rightleftharpoons  \pi \Omega,            \\
        N \bar{N}      &~\text{annihilation},                      &\quad 
        N \bar{Y}      &~\text{annihilation},                      \\
        J/\psi + N     &\rightleftharpoons  Y_c + \bar{D},         &\quad
        \psi' + N      &\rightleftharpoons  Y_c + \bar{D},         \\
        J/\psi + \pi   &\rightleftharpoons  D + \bar{D^*},         &\quad
        \psi' + \pi    &\rightleftharpoons  D + \bar{D^*},         \\
        J/\psi + \rho  &\rightleftharpoons  D + \bar{D},           &\quad
        \psi' + \rho   &\rightleftharpoons  D + \bar{D},           \\
        D + \pi        &\rightleftharpoons  D^* + \rho,            &\quad
        D^* + \pi      &\rightleftharpoons  D + \rho,              \\
        D + N          &\rightleftharpoons  D^* + N,               &\quad
        D^* + \bar{N}  &\rightleftharpoons  D + \bar{N},
    \end{aligned}
\end{equation*}
where $ Y $ stands for $ \Lambda $ or $ \Sigma $, and $ Y_c $ stands for 
$ \Lambda_c $ or $ \Sigma_c $.

The experimental data inspired parameterization of total cross sections are 
adopted~\cite{Koch:1986ud,Schopper:1988hwx}. For the channels with missing 
data, the total cross sections are assumed as constant values and/or calculated 
using the additive quark model 
(AQM)~\cite{Bierlich:2022pfr,Levin:1965mi,Lipkin:1973nt}
\begin{equation}
    \begin{aligned}
    \sigma_{AQM}^{h_1 h_2} &= (40~\text{mb}) 
                             \frac{n_{eff}^{h_1}}{3} \frac{n_{eff}^{h_2}}{3}, \\
    n_{eff} &= n_d + n_u + 0.6 n_s + 0.2 n_c + 0.07 n_b.
    \end{aligned}
    \label{eq:AQM}
\end{equation}
In Eq.~(\ref{eq:AQM}), $ h $ denotes the hadron. $ n_{eff} $ and $ n_i $ refer to 
the total effective valence quark number and the effective $i$-flavor valence quark number 
in a hadron, respectively.

\subsubsection{Hadron decay} \label{subsubsec:hadron_decay}
By default, only the particles with the mean proper lifetime greater than 
1 cm/$ c = 10^{13} $ fm/$c$~\cite{ALICE:2017hcy} are recognized as stable 
in PACIAE 4.0. For simplification, we assume that all the unstable 
particles produced through the hadronic rescattering process will survive 
until the end of the hadronic rescattering. Of course, it is easy to switch on 
the decay of unstable particles in the transport process. So far, the decay of 
the unstable particles is handled by PYTHIA 6~\cite{Sjostrand:2006za}. In the 
next version, we will plug in the PYTHIA 8 decayer~\cite{Bierlich:2022pfr}.

\section{Summary} \label{sec:summary}
We have upgraded the Parton And-hadron China Institute Atomic Energy (PACIAE) 
model to a new version PACIAE 4.0, which is already available to the public. 
The program has been refactored and 
rewritten by the free-format modern Fortran mixed with C++ languages instead of 
the fixed-format FORTRAN 77. This is the first version of PACIAE model 
transformed from Fortran to C++. The interface to PYTHIA 8.3 has been provided. 
PACIAE 4.0 now can connect to both Fortran-based PYTHIA 6.4 and 
C++-based PYTHIA 8.3. Several improvements have also been introduced to model 
the high-energy collisions.

\section*{Acknowledgements} \label{sec:acknowledgements}
This work is supported by the National Natural Science Foundation of China 
under grant Nos. 12375135, 11447024, and 11505108. A.K.L. acknowledges the 
financial support from the China Scholarship Council. Y.L.Y. acknowledges the 
financial support from the Continuous-Support Basic Scientific Research Project 
in China Institute of Atomic Energy. L.Z. acknowledges the support from the 
Fundamental Research Funds for the Central Universities, China University of 
Geosciences (Wuhan) with No. G1323523064. W.C.Z. is supported by 
the Natural Science Basic Research Plan in Shaanxi Province of China 
(No. 2023-JCYB-012). H.Z. acknowledges the financial support from Key 
Laboratory of Quark and Lepton Physics in Central China Normal
University under grant No. QLPL2024P01.


\end{document}